%% file: ms_final.tex
\documentclass[structabstract]{aa}
\usepackage{graphicx}
\usepackage{amsmath}
\usepackage{amssymb}
\usepackage{epsfig} 
\usepackage{txfonts}
\usepackage{color}
\usepackage{xcolor}
\usepackage{url}
\usepackage{natbib}
\bibpunct[]{(}{)}{;}{a}{}{,}

\newcommand{\otto}{KIC~7341231}
\newcommand{\bob}{HD~52265}

\begin{document}

\title{Asteroseismic inversions for radial differential rotation of Sun-like stars: Sensitivity to uncertainties}
\titlerunning{Sensitivity of asteroseismic inversions to stellar models}

\author{
H.~Schunker \inst{\ref{inst1}}
\and
J.~Schou \inst{\ref{inst1}}
\and
W.~H.~Ball \inst{\ref{inst2}}
}

\institute{
Max-Planck-Institut f\"{u}r Sonnensystemforschung, Justus-von-Liebig-Weg 3, 37077  G\"{o}ttingen, Germany \label{inst1}\\
\email{schunker@mps.mpg.de}
\and
Georg-August-Universit\"{a}t G\"{o}ttingen, Institut f\"{u}r Astrophysik, Friedrich-Hund-Platz 1, 37077   G\"{o}ttingen, Germany\label{inst2}\\
}

\date{Received $\langle$date$\rangle$ / Accepted $\langle$date$\rangle$}

\abstract
{}
{We quantify the effect of observational spectroscopic and asteroseismic uncertainties on regularised least squares (RLS) inversions for the  radial differential rotation of Sun-like and subgiant  stars.}
{
We first solved the forward problem to model rotational splittings plus the observed uncertainties for models of a Sun-like star, \bob, and a subgiant star, \otto.
We randomly perturbed the parameters of the stellar models within the uncertainties of the  spectroscopic and asteroseismic constraints and used these perturbed stellar models to compute rotational splittings.
We experimented with three rotation profiles: solid body rotation, a step function, and a smooth rotation profile decreasing with radius.
We then solved the inverse problem to infer the radial differential rotation profile using a RLS  inversion and kernels from the best-fit stellar model.
We also compared RLS, optimally localised average (OLA)  and direct functional fitting inversion techniques.
}
{
We found that the inversions for Sun-like stars with solar-like radial differential rotation profiles  are insensitive to the uncertainties in the stellar models.  The uncertainties in the splittings dominate the uncertainties in the inversions and solid body rotation is not excluded.
We found that when the rotation rate below the convection zone is increased to six times that of the surface rotation rate the inferred rotation profile excluded solid body rotation. 
We showed that when we reduced the uncertainties in the splittings by a factor of about 100, the inversion is sensitive to the uncertainties in the stellar model.
With the current observational uncertainties, we found that inversions of subgiant stars are sensitive to the uncertainties in the stellar model.
}
{
Our findings suggest that inversions for the radial differential rotation of subgiant stars would benefit from more tightly constrained stellar models.
We conclude that current observational uncertainties make it difficult to infer radially resolved features of the rotation profile in a Sun-like star using inversions with regularisation.
In Sun-like stars, the  insensitivity of the inversions to stellar model uncertainties suggests that it may be possible to perform ensemble inversions for the average radial differential rotation of many stars with a range of stellar types to better constrain the inversions. 
}

\maketitle


\section{Introduction}\label{intro}

The differential rotation of the solar interior is an important constraint for solar dynamo models \citep{Charbonneau2010}. However, a successful solar dynamo model is yet to be developed. Studying the interior rotation of other Sun-like stars helps to understand the relationship between differential rotation and activity in our Sun.

The interior rotation of a star can be inferred by perturbations to the frequencies of its acoustic oscillation modes. This perturbation is referred to as rotational splitting and is proportional to the rate of rotation. 
Inferring the  radial differential rotation of stars requires measuring the splittings of modes and solving an inversion problem for the rotation using some stellar model.
The splitting of oscillation modes by rotation in solar-like oscillators with solar-like rotation profiles is typically seen as a broadening rather than a distinct separation of the mode peaks in the power spectrum caused by the combined effect of the mode linewidths (lifetimes of the oscillations) and the slow rotation rates. In subgiants, the magnitude of rotation is larger and therefore the splittings are more easily measured.

As a result of space-based observations from missions like CoRoT \citep{Baglinetal2006} and \textit{Kepler} \citep{Boruckietal2010} measuring the mode splittings has recently become possible.
Splittings that clearly vary with radial order have been observed in subgiants \citep{Deheuvelsetal2012,Deheuvelsetal2014} and white dwarfs \citep{Kawaler1999}. 
In these stars, the core to surface gradient of rotation is large and the splitting of modes that are sensitive to the core conditions can be easily observed.
However, the radial differential rotation profiles of subgiant stars inferred from  different inversion methods are not in agreement. In addition, they include  scenarios that are not intuitive, where the core and envelope counter-rotate \citep{Deheuvelsetal2012,Deheuvelsetal2014}. 
Rotational splittings of Sun-like stars measured as a function of frequency have constrained the radial differential rotation to be consistent with the Sun  \citep{Nielsenetal2014}, but inversions for these stars have not yet been attempted. 

Inverting for the rotation given the rotational splittings is, mathematically, an ill-posed problem. It is usually solved using some flavour of linear inversion technique, such as regularised least squares (RLS), which have proved successful for the case of the Sun \citep[e.g.][]{JCDST1990}. Another way is to invert onto fixed functional bases \citep[e.g.][]{Kawaler1999}. Both of these techniques rely on having a good stellar model to describe the sensitivity of modes to rotation at different radii, called the  rotation kernels.
Observational uncertainties enter into any given inversion technique both from the asteroseismic rotational splittings and constraints on the stellar model: the oscillation frequencies  and  spectroscopic constraints  (effective temperature, surface gravity, and metallicity).
The uncertainty in the constraints is due to the precision of the stellar model, and thus the rotation kernels.
While solar models are well constrained, the uncertainty in stellar models is much larger. The sensitivity of the inversion procedure to the uncertainty in the stellar models has not  previously been quantified.

\citet{Deheuvelsetal2014} inferred the radial differential rotation  for a subgiant star with two stellar models computed by different stellar evolution codes (ASTEC and CESAM2K). They found no significant difference in the inversion results. However, best-fit models from these two stellar evolution codes are consistent to first order \citep{Monteiroetal2006}. 
In this paper, we explore\ the difference in the stellar models due to non-seismic uncertainties in the larger constraints.

To accomplish this, we  quantify the sensitivity of the RLS inversions for radial differential rotation to the use of rotation kernels constructed from an improper stellar model for a Sun-like star, \bob, and a subgiant star, \otto.
Section~\ref{sect:method} describes our experiment to test the effect of uncertainties on the inversions, and  the RLS inversion technique is described in Sect.~\ref{sect:RLS}. 
Section~\ref{sect:resultsotto} then illustrates the results for the subgiant star \otto, followed by the Sun-like star \bob \, in Sect.~\ref{sect:resultsbob}. We then explore a simpler inversion technique, which returns the sign of the gradient of the inverted rotation profile in Sect.~\ref{sect:funcfit}. We discuss the implications of the results towards establishing which quantities of the radial differential rotation  can be reliably determined in Sect.~\ref{sect:conc}.

\section{Method to test sensitivity of inversions to observational uncertainties}\label{sect:method}
First we solved the forward problem. From the observed spectroscopic and asteroseismic parameters of the star, we computed a best-fit stellar model, which we designate as the  {reference model}. We then randomly perturbed the observed parameters, both spectroscopic and asteroseismic, within the uncertainties and computed best-fit models for those parameters, which we designate as the {perturbed model}. 

We constructed three synthetic radial differential rotation profiles, a solid body, a step function, and a smooth function, with amplitudes based on the measured bulk rotation rates of \bob \, and \otto. In  stars of both these types, we expect that the core  rotates faster than the convective envelope and that the rotation may change most dramatically at the base of the convection zone (BCZ), so all of the rotation rate profiles decrease as a function of radius.

For a given synthetic radial differential rotation profile coupled with a stellar model, we compute a set of rotational splittings and add a model of the uncertainties  based on the observed rotational splitting uncertainties.
The uncertainties discussed in this paper are: 
\begin{itemize}
\item $\sigma_i(\delta \omega)$ is the  uncertainty in the observed rotational splitting of the $i$th mode, where $i$ indicates the radial order and harmonic degree of the mode $i=(n,\ell)$; 
\item $\sigma_i(\omega)$ is the  uncertainty in the observed frequency of the mode;
\item $\sigma_k(\mathrm{obs})$ is the uncertainty on the observational constraints on the stellar model, including $\sigma_i(\omega)$ in addition to the spectroscopic constraints;
\item $\sigma_j(\overline{\Omega})$ is the uncertainty in the inverted rotation profile at each radial point, $j$ (see Eq.~\ref{eqn:stddev}).
\end{itemize}

\section{Least squares inversion with smooth regularisation} \label{sect:RLS}

In the Sun, the RLS inversion method with a smoothness condition (Tikhonov regularisation)  is used successfully to constrain the interior rotation profile in two dimensions.
We choose to use the RLS inversion since it has been used in attempts to invert for  the radial differential rotation  of other stars and  is representative of the general inversion techniques with regularisation.

We follow the procedure of \citet{JCDST1990} and minimise 
\begin{equation}
\sum_{i=1}^M \left[ \delta \omega_i - \sum_{j=1}^N \overline{\Omega}(r_j) \int_0^R K_{\mathrm{ref},i}(r) \phi_j (r) dr \right]^2 + \mu F(\overline{\Omega})
,\end{equation}\label{eqn:tomin}
where $\overline{\Omega}$ is the inverted rotation profile, $\phi_j(r)$ are a chosen set of basis functions, $\mu$ is the trade-off parameter (discussed in Sect.~\ref{sect:tradeoff}), $K_{\mathrm{ref},i}$ is a kernel from the reference stellar model, and $F(\overline{\Omega})=\int_0^R |d\overline{\Omega}/dr|^2 dr$ is the regularisation function to ensure the result is smooth.
The basis functions are 
\begin{equation}
\phi_j = \left\{ 
                \begin{array}{l l}
                      1, & \quad x_{j-1} < r \le x_j  \\
                      0, & \quad \mathrm{elsewhere} 
               \end{array} \right.
,\end{equation}
where  $j=1,...,N$ and $x_0=0, $ and $x_N=R$. We solve on a uniform grid of 200 equidistant points $r_j = (x_{j-1} + x_j)/2$. The results do not change significantly if this number of points is halved or doubled. 
The inverted rotation profile at point $j$ is given by 
\begin{equation}
\overline{\Omega}(r_j) = \sum\limits_{i=1}^M c_{i}(r_j)\delta \omega_i
,\end{equation}
where $c_{i}$ are the inversion coefficients.

The standard deviation at each point $j$ of the resulting inversion profile is given by
\begin{equation}
\sigma_j(\overline{\Omega})=\sqrt{ \sum\limits_{i=1}^M \left[ \sigma_i(\delta \omega) c_i(r_j) \right] ^2 }.
\end{equation}\label{eqn:stddev}

\subsection{Trade-off parameter}\label{sect:tradeoff}

The choice of the trade-off parameter $\mu$ is non-trivial. One way to do this is to compute  inversions for a wide range of $\mu$ and to find the point of maximum curvature in the resulting L-curve, which is a measure of the smoothness of the solution versus the agreement with the observed data.
For stellar rotation inversions, the choice is generally more subjective than using the L-curve. The trade-off parameter is selected based on what the observer expects the inverted rotation profile to look like and the agreement with the observed rotational splittings \citep[e.g.][]{Deheuvelsetal2012}.

We tuned the trade-off parameter for \bob \, by incrementing $\mu$ from $10^{-15}$ to $10^{20}$ in powers of 10 to find the best fit to the synthetic rotation profile. 
Having the luxury of knowing the input rotation profile, we selected $\mu=10^7$ as a good fit. For \otto, we found that the values used by \cite{Deheuvelsetal2012} returned a good approximation to the synthetic input rotation, and so we used the same normalisation, $u\mu$, consisting of two parameters where $\mu=10^{-4}$ and $u=1/\langle \sigma_i(\delta \omega)^2 \rangle$. 
For \bob, \, we chose $\mu=10^{7}$.

\begin{center}
\begin{figure}
\includegraphics[width=0.5\textwidth]{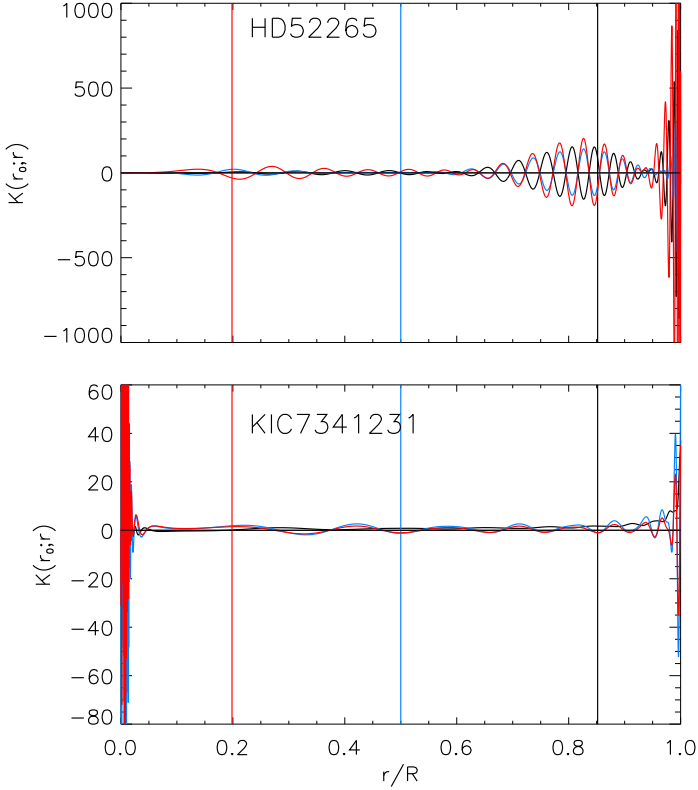}
\caption{Example averaging kernels for the RLS inversions for \bob \, with $\mu=10^{7}$, $16\le n \le25$, and $\ell=1,2$ (top) and \otto \, $\mu=10^{-4}$, $\ell=1$ modes in the range $285 \le \nu \le 485$~nHz (bottom). The target depths for each averaging kernel are indicated with the vertical line with the corresponding colour. None of the kernels are well localised.}
\label{fig:avek}
\end{figure}
\end{center}

Figure~\ref{fig:avek} shows the averaging kernels are not well localised at the target depths and are predominantly sensitive more towards  the surface.
The averaging kernels, $K(r_0;r)=\sum_{i}c_i(r_0)K_i(r)$, show the sensitivity of the rotation measurement at one depth $r_0$ for a given inversion method and regularisation.
Well localised rotation kernels would have a peak at the target depth \citep[examples of well-localised rotation kernels for the Sun incorporating many more modes can be found in Fig.~13 of ][]{JCDST1990}. Figure~\ref{fig:avek} shows that the kernels of \bob \, are not well localised and are sensitive to similar depths. 

\section{Subgiant stars: Sensitivity of RLS inversions to observational uncertainties }\label{sect:resultsotto}
In this section, we quanitfy the sensitivity of the inversions to uncertainties in the splittings, $\sigma_i(\delta \omega)$, and uncertainties in the stellar model constraints, $\sigma_k(\mathrm{obs})$, for an subgiant star \otto.

\subsection{Stellar model}\label{sect:ottomodel}

We computed stellar models for \otto \, in the same way as \citet{BallGizon2014} with the Modules for Experiments in Stellar Astrophysics code \citep[MESA\footnote{\url{http://mesa.sourceforge.net/}} revision 6022;][]{Paxton2013}. The spectroscopic parameters reported in \cite{Deheuvelsetal2012}, and the asteroseismic frequencies from $\sim9$~months of observations  reported in \citet{Appx2012} were used to constrain the stellar models.  
Our sample of stellar models contains one designated reference model minimised with respect to the observed spectroscopic and asteroseismic constraints as well as 20 additional so-called perturbed models fit to random realisations of the observational constraints perturbed within their uncertainties (see Table~\ref{tab:perts} for the specific parameters).
Here, we summarise the fitting procedure and additional details are found in the original paper \citep{BallGizon2014}.
\begin{table}
\caption{Constraints and uncertainties for the spectroscopic parameters of \bob \, from \citet{Ballotetal2011} and \otto \, from \citet{Deheuvelsetal2012}.}
\centering
\begin{tabular}{lcccc}
Star &   $T_\mathrm{eff}$  & $\log~g$ & $\mathrm[Fe/H]$ & $\langle \sigma_i(\omega) \rangle$ \\
     &    [K]             &  & $\mathrm{[dex]}$ &  [$\mu$Hz]\\
\hline
\otto   & $5233 \pm 100$ & $3.55 \pm 0.03$ & $-1.64 \pm 0.10$  & 0.11 \\
\bob    & $6100 \pm 60$  & $4.35 \pm 0.09$ & $0.19 \pm 0.05$ & 0.12 \\
\hline
\end{tabular}
\label{tab:perts}
\end{table}

Each model was fit to both seismic and non-seismic constraints, by optimizing the total $\chi^2$ with respect to the stellar model's age, mass, initial metallicity, initial helium abundance, and mixing-length parameter. That is, we minimise
\begin{equation}
\chi_\mathrm{mod}^2 = \sum\limits_{k=1}^{N} \left(  \frac{x_{k,\mathrm{mod}} - x_{k,\mathrm{obs}} }{\sigma_k(\mathrm{obs})}
\right)^2
\label{eqn:chi2}
,\end{equation}
where $x_{k,\mathrm{obs}}$ represent all of the $N$ seismic and non-seismic constraints, and $\sigma_k(\mathrm{obs})$ their corresponding uncertainties, and $x_{k,\mathrm{mod}}$ are the values of the stellar model. The non-seismic constraints are the effective temperature $T_\mathrm{eff}$, surface gravity $\log g$, metallicity $\mathrm{[Fe/H]}$, and luminosity $\log L_*$, with the former determined from high-resolution spectroscopy and the latter from Hipparcos observations. The seismic constraints were all of the 40  $\ell=0,1,2,3$ modes that were detected and correctly fit as listed in Table A.17 in \citet{Appx2012}.
The surface effects, known systematic discrepancies between observed and modelled frequencies of high-order modes, were fit using a function of the form $\nu^3/\mathcal{I}$, where $\nu$ and $\mathcal{I}$ are a mode's frequency and inertia normalised at the photosphere \citep{BallGizon2014}. Mode frequencies and inertiae were computed with the Aarhus adiabatic oscillation package \citep[ADIPLS; ][]{ADIPLS}. These stellar models provide us with the rotation kernels.

\subsection{Synthetic rotation profile and mode splittings}\label{sect:rotprofs}
We consider three rotation profiles (see Fig.~\ref{fig:rotprofotto}):
   \enumerate{
   
   \item Solid body: 
   \begin{equation}
        \hspace{-3cm}
   \Omega(r) =\Omega_\mathrm{c}
   \end{equation}

   \item Smooth function: 
    \begin{equation} 
    \Omega(r) = \Omega_\mathrm{c} \left[ (h -1) \gamma + 1 \right] 
    ,\end{equation}\label{eqn:smooth}
   \quad where   $h=\left( \cos(r \pi/R) + 1 \right)/2$.
   
  \item Step function: 
  \begin{equation}
     \Omega(r) =\left\{ 
     \begin{array}{l}
         \Omega_\mathrm{c} \quad 0 < r/R \le r_\mathrm{BCZ}\\     
         \Omega_\mathrm{e} \quad \text{$r_\mathrm{BCZ} < r/R \le 1.0$}        
      \end{array} \right.
  ,\end{equation} 
where $R$ is the stellar radius,    $\Omega_\mathrm{c}$ is the core angular velocity, $\Omega_\mathrm{e}$ is the angular velocity of the envelope, and $r_\mathrm{BCZ}$ is the radius at the base of the convection zone. 
The values of the parameters for each star are given in Table~\ref{tab:synthrotpar} and the rotation profiles are shown in Fig.~\ref{fig:rotprofotto}.

\begin{table}
\caption{Characteristics of the  synthetic rotation profiles for stars \bob \, and \otto. \label{tab:models}}
\centering
\begin{tabular}{lcccc}
Star &   $\Omega_c/2\pi$  & $\Omega_e/2\pi$ & $\gamma$ & $r_\mathrm{BCZ}$  \\
    &   [nHz]  & [nHz] &  & [R]  \\
\hline
\bob      & 1035  & 685 & 0.35 & 0.8   \\
\otto   & 700 & 209 & 0.72 & 0.2  \\
\hline
\end{tabular}
\label{tab:synthrotpar}
\end{table}

For a spherically symmetric, non-rotating star the frequencies of the oscillation modes, $\omega$, are functions of radial order, $n$, and harmonic degree, $\ell$, but independent of azimuthal order, $m$. 
When the rotation rate is slow, the frequency is perturbed so that
$\omega = \omega_{n\ell} + m \delta \omega_{n\ell}$, ignoring any latitudinal differential rotation. 
The rotational splittings are a linear function of the angular rotation rate, $\delta \omega_{n\ell } = \int_0^R K_{n\ell}(r)\Omega(r) dr = 2 \pi a_1(n,\ell)$, where $K_{n\ell}$ are the rotational kernels of the star and $a_1$ are the expansion coefficients for the splitting  \citep{Schouetal1998}.

\begin{center}
\begin{figure}
\includegraphics[width=0.45\textwidth]{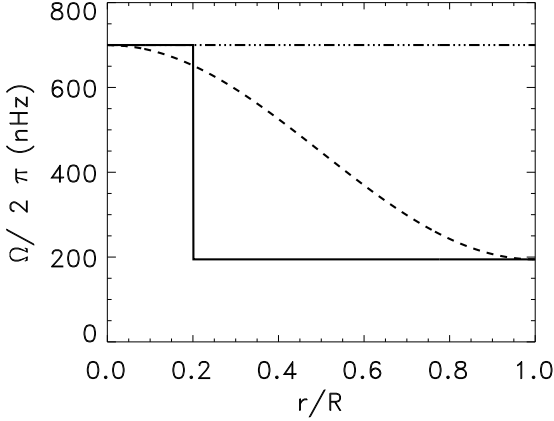}
\caption{Three synthetic rotation profiles described in Sect.~\ref{sect:rotprofs}: solid-body (dash-dot), step-function (solid), and smooth function (dashed).  The step function discontinuity lies at the base of the convection zone. The magnitudes and sign of the gradient are based on previous estimates of the rotation rate by \citet{Deheuvelsetal2012}.
}
\label{fig:rotprofotto}
\end{figure}
\end{center}

\subsection{Noise model for the rotational splittings}\label{sect:noiseotto}

For \otto,\,we used the measured mode frequency uncertainties  for $\ell=1,2$ from nine months of \textit{Kepler} observations of \otto \citep{Appx2012}. We point out that the uncertainties reduce by the square root of the number of observations and we use this dataset as a model, which can be extrapolated to longer data series.

Using the relationship $a_1 \approx \delta \omega_{n\ell m}/m$ and assuming that the uncertainty in the frequency measurement for each azimuthal order $m$ is constant, the uncertainties in the splitting coefficients are given by 
$\sigma_i(\delta \omega)=\sigma_i(\omega)/\sqrt{\frac{1}{3} \ell(\ell+1)}$.
Our model for the noise is a quadratic function fit to the uncertainties of the splittings, $\sigma_i(\delta \omega)$, for each $\ell$ as a function of frequency  (see Fig.~\ref{fig:splitnoiseotto}). 

\begin{center}
\begin{figure}
\includegraphics[width=0.5\textwidth]{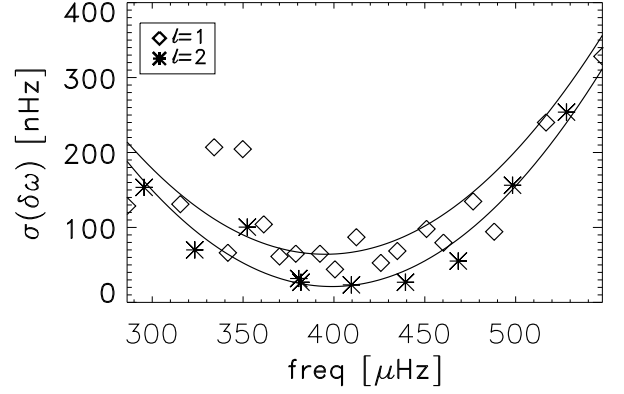}
\caption{
Models of uncertainties for the synthetic rotational splittings used for \otto \, for the different modes, $\ell=1,2$ for 9 months of \textit{Kepler} data from \citet{Appx2012}.
The solid curves are a quadratic fit for each $\ell=1,2$ which we use to model the noise for the synthetic rotational splittings.
}
\label{fig:splitnoiseotto}
\end{figure}
\end{center}

The mean of the modelled uncertainties on the splittings for \otto \, is $\langle \sigma_i(\delta \omega) \rangle = 93$~nHz.
We add a Gaussian distributed random realisation of the noise centred at zero and standard deviation given by the modelled quadratic fit to the splittings.

\subsection{Inversion results: Uncertainties in both the stellar model constraints and the rotational splittings}\label{sect:noisecompotto}

We now solve the inverse problem for each set of  splittings using the rotation kernel derived from the reference model to get what we will refer to as perturbed rotation profiles. 
In this case, we are using so-called improper kernels to solve the inversion, since they are not consistent with what was used in the forward problem. 

In accordance with what has been observed for this star  \citep{Deheuvelsetal2012} we use rotational splittings in the frequency range $285 < \nu < 485$~nHz and with harmonic degree $\ell=1$ for \otto.

Figure~\ref{fig:ell12otto} shows the  inversion results for \otto \, computed using each instance of improper rotation kernel as well as one identical realisation of noise on each set of splittings. 
The reference rotation profile is the inverted rotation profile solved using the reference rotation kernels (derived from the reference stellar model) to compute both the splittings and the inversion (red curves in Fig.~\ref{fig:ell12otto}).

\begin{table}
\caption{ Dispersion ratio (Eq.~\ref{eqn:disp}) of the inversions for rotational splittings computed with perturbed stellar models of \otto.}  
\vspace{-0.5cm}
\label{table:dispotto} 
\centering 
\include{otto_table}
\tablefoot{
Frequency range for the first three columns is $285~\mu\mathrm{Hz} \le \nu \le 480~\mu$Hz (22 modes). 
The first column includes only $\ell=1$ modes and the rest  $\ell=1,2$ modes.
In the third column, we reduced the uncertainty in the splittings to the equivalent of three years of observations. 
In the last two columns, we increased the frequency range to $200~\mu\mathrm{Hz} \le \nu \le 530~\mu$Hz (26 modes) and decreased the frequency range to $350~\mu\mathrm{Hz} \le \nu \le 450~\mu$Hz (13 modes) respectively, changing the uncertainties. 
}
\end{table}

To quantify the sensitivity, we integrated the uncertainty in the inversion and divided by the standard deviation of the differences between the perturbed inversion results and the reference inversion result, 
\begin{equation}
\mathcal{D}  = \frac{ \int_0^R \sigma \left( \overline{\Omega} \right)\mathrm{dr}}{ \mathrm{stddev}\left\{ \int_0^R | \overline{\Omega}_{\mathrm{pert},k} - \overline{\Omega}_\mathrm{ref} | \mathrm{dr} \right\} }.
\label{eqn:disp}
\end{equation}
We refer to this as the dispersion ratio, where $k$ is each of the stellar models in the sample (see Sect.~\ref{sect:ottomodel}).

When $\mathcal{D} \approx 1$, the spread of the inversion results due to using a rotation kernel derived from an improper stellar model is of the order of the uncertainty in the noise on the observed rotational splittings. 
When $\mathcal{D} \gg 1$, the inversion results are dominated by the noise in the rotational splittings, and when $\mathcal{D} \ll 1$, the inversion results are most sensitive to the use of the rotation kernels derived from an improper stellar model.

\begin{center}
\begin{figure}
\includegraphics[width=0.45\textwidth]{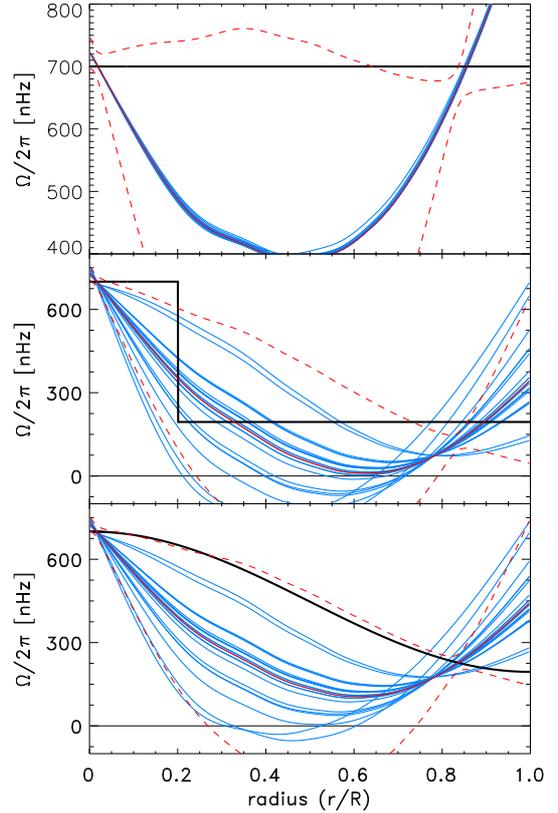} 
\caption{
Inverted rotation profiles for the perturbed models of \otto \, (blue curves) with one realisation of the rotational splitting noise using $\ell=1,2$ splittings in the range $285 \le \nu \le 480~\mu$~Hz. 
The top panel shows the inversion results for solid body rotation, the middle panel for a step function profile, and the bottom panel for the smooth profile (all shown in solid black curves).
The reference inversion (using the reference kernels for the forward and inverse problem) is shown in red, and the standard deviation (Eq.~\ref{eqn:stddev}) is the dashed red curve.
}
\label{fig:ell12otto}
\end{figure}
\end{center}

We experimented with varying the range of frequency, number of modes, noise level, and harmonic degree to compute the dispersion ratio (see Table~\ref{table:dispotto}).
In the case of solid body rotation, the inversion is consistently dominated by noise on the rotational splittings.
Using $\ell=1,2$ halves the dispersion ratio compared to using only $\ell=1$ modes (compare Column~1 to Column~2, and Fig.~\ref{fig:ell12otto} corresponds to Column~2). 
Table~\ref{table:dispotto} also shows that using both $\ell=1,2$ modes and a full three years of observations (Column~3), the dispersion ratio becomes smaller for the step function and smooth synthetic rotation. 
Increasing the frequency range and subsequently the number of observed modes (Column~4)  decreases the dispersion ratio so that $\mathcal{D}=1$ and one must be careful to use the correct stellar model when inverting for the rotation. 
Using both $\ell=1,2$ modes with the wider range of frequency and hence more modes, despite the larger uncertainties on those splittings, gives a dispersion ratio closest to one.


\section{Sun-like stars: Sensitivity of RLS inversions to observational uncertainties}\label{sect:resultsbob}
We repeat the experiment for a solar-like star, \bob, for which it is much harder to measure the rotational splittings due to the mode linewidths, low rotation rates, and purely acoustic nature of the oscillations (no mixed modes) with similar depth sensitivities.

\subsection{Stellar models}
We used the stellar models computed for \bob \, by \citet{BallGizon2014} with the MESA code, as described in Sect.~\ref{sect:ottomodel}. 
The sample contains one {reference} model minimised with respect to the observed spectroscopic and asteroseismic constraints from \citet{Ballotetal2011} as well as 100 additional {perturbed} models fit to random realisations of the observational constraints perturbed within their uncertainties (see Table~\ref{tab:perts} for the specific parameters).
For \bob, the seismic constraints were oscillation frequencies for ten radial, ten dipole, and eight quadrupole modes as reported by \citet{Ballotetal2011}. We omitted the lowest order modes of each degree, which were noted in \cite{Ballotetal2011} as less reliable.

\subsection{Synthetic rotation profile and mode splittings}
Previous measurements of the rotation of \bob \, give an average rotation rate of $\Omega/2\pi=980$~nHz, about 2.3 times that of the Sun \citep[][ and references therein]{Gizonetal2013}, therefore, the amplitude of the synthetic profiles we created are in this range. 
Most stellar models place the convection zone at close to $r_\mathrm{BCZ}=0.8$~R. 
Figure~\ref{fig:rotprofbob} shows the synthetic rotation profiles, with equations for the rotation given in Sect.~\ref{sect:rotprofs} and parameters in Table~\ref{tab:synthrotpar}.
\begin{center}
\begin{figure}
\includegraphics[width=0.45\textwidth]{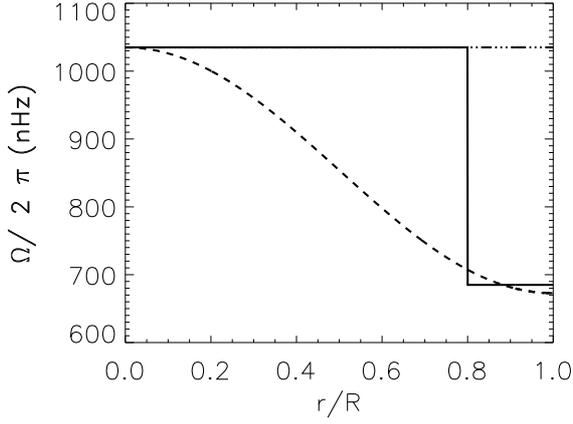}
\caption{Synthetic rotation profiles for \bob. 
The magnitude and sign of the gradient are based on previous estimates of the rotation rate by \citet{Gizonetal2013}.
}
\label{fig:rotprofbob}
\end{figure}
\end{center}

\subsection{Noise model for the rotational splittings}
\begin{center}
\begin{figure}
\includegraphics[width=0.5\textwidth]{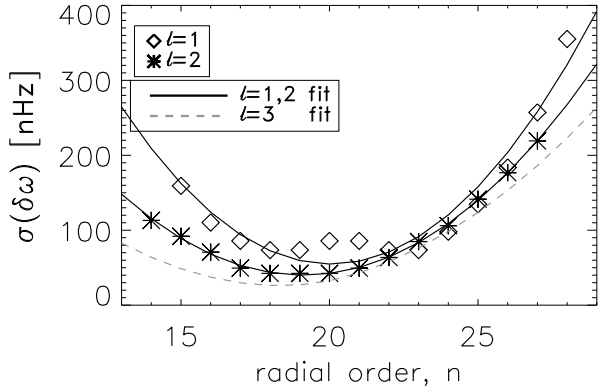}
\caption{Uncertainties in the rotational splitting measurements for \bob.
The solid curves are a quadratic fit for each $\ell=1,2,$ which we use to model the noise for the synthetic rotational splittings.
The dashed curve is the fit for $\ell=2$ scaled by the ratio between the fit for $\ell=2$ and $\ell=1,$ which we use as the model of uncertainty in the splittings for $\ell=3$. 
}
\label{fig:splitnoisebob}
\end{figure}
\end{center}
We scale the measured frequency uncertainties for four months of VIRGO solar data from \citet{Stahn2011} by $1/\sqrt{\frac{1}{3} \ell(\ell+1)}$ to model the uncertainties of the splittings.
We  fit a quadratic function to the uncertainties of the splittings, $\sigma_i(\delta \omega)$, as a function of radial order for each $\ell=1$ ($n=15,...,28$) and $\ell=2$ ($n=4,...,27$) independently (see Fig.~\ref{fig:splitnoisebob}).
The mean of the modelled uncertainties on the splittings for \bob \, is $\langle \sigma_i(\delta \omega)\rangle = 80$~nHz, which is of the order of the observed uncertainty on the bulk rotation rate of \bob  \, \citep[30~nHz ][]{Gizonetal2013}.
We add Gaussian distributed random realisation of noise in the same way as described in Sect.~\ref{sect:noiseotto}.

\subsection{Inversion results: Uncertainties in both the stellar model constraints and the rotational splittings}
Figure~\ref{fig:rlsmodels} shows the rotation inversion results for \bob \, computed with each instance of improper rotation kernels as well as one identical realisation of noise on each set of splittings. The inversion is completely dominated by the noise on the rotational splittings, quantified by the dispersion ratio, $\mathcal{D}\approx 100$ (Table~\ref{table:dispbob100}, Column~1).

\begin{center}
\begin{figure}
\includegraphics[width=0.5\textwidth]{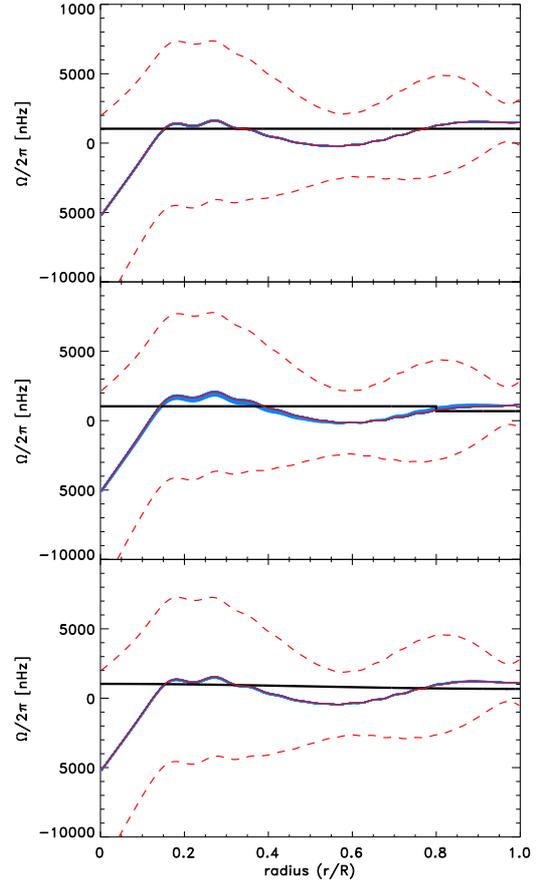}
\caption{
Inverted rotation profiles for the 100 perturbed models of \bob \, (blue curves, very close to the red curve) with one realisation of the rotational splitting noise. 
The top panel shows the inversion results for solid body rotation, the middle panel for a step function profile, and the bottom panel for the smooth profile (all shown in solid black curves).
The inverted reference rotation profile (using the reference kernels for the forward and inverse problem) is shown in red, and the standard deviation (Eq.~\ref{eqn:stddev}) are the dashed red curves.
}
\label{fig:rlsmodels}
\end{figure}
\end{center}

We considered larger ($10 \le n \le 25$) and smaller ($18 \le n \le 22$) ranges of modes in radial order, and we also add the $\ell=3$ mode (with $n=16,...,25$) rotational splittings.
 Table~\ref{table:dispbob100} shows the dispersion ratio values for these cases.
Including $\ell=3$ modes reduces the uncertainty to $\langle \sigma_i(\delta \omega) \rangle = 72$~nHz.
In all cases, the dispersion ratio remains of the order of 100, dominated by the uncertainties in the rotational splittings. 
The inversions are completely insensitive to the stellar model used.

This analysis can be extended to longer more realistic time series. Since the stochastic noise decreases with time, the uncertainties decrease with the square root of the number of observations. For Kepler, there would be nine times as many observations, and therefore the uncertainties would be reduced by three. This is not significant enough to matter. The uncertainties in the stellar model begin to matter when the uncertainties are reduced by an order of 100, which requires 10,000 times as much data (see Table~\ref{table:dispbob100}, column 3).

\begin{table}
\caption{Dispersion ratios (Eq.~\ref{eqn:disp}) of the inversions for rotational splittings computed with perturbed stellar models of \bob. 
}  
\vspace{-0.5cm}
\label{table:dispbob100} 
\centering 
\include{bob100_table}
\tablefoot{
First column shows the dispersion ratio for the nominal case with typical observed uncertainties on the rotational splittings and range of modes.
The second and third columns have the observed uncertainty in the splittings reduced by factors of 10 and 100.
The fourth and fifth column have the number of modes increased and decreased, respectively, by changing the range of the radial orders, which changes the uncertainty.
The last column shows the dispersion ratio for inversions including  $\ell=1,2,3$ modes, subsequently the uncertainty is also reduced.
The trade-off parameter is always $\mu=10^{7}$.}
\end{table}


\subsection{Increasing the core-to-surface rotation gradient}\label{sect:bigratios}
In the Sun, the core-to-surface rotation gradient at mid-latitudes is $\approx 100$~nHz \citep{Schouetal1998}, of the same order as we have given \bob.
Not all stars with solar-like oscillations  necessarily have a Sun-like rotation profile, as we have assumed here.
Stars generally spin down with age \citep{Skumanich1972}, so younger stars  rotate faster and have more scope for larger radial differential rotation gradients.
We repeated the experiment for \bob \  with our smooth rotation profile, but with increasing core rotation rates. Specifically, we multiplied the first term in Eq.~\ref{eqn:smooth} by a factor $\alpha=5, 10, 15, 20, \, \textrm{and} \, 25$. We used the standard level of noise on the splittings, RLS regularisation parameter $\mu=10^{7}$, and modes with radial order $n=16,...,25$ and harmonic degree $\ell=1,2$.

Figure~\ref{fig:refbob} shows that the dispersion ratio  approaches unity  as the core-to-surface rotation  gradient increases. 
As an additional measure of the difference in the results we have plotted the integrated difference between the inverted rotation profile and the model rotation profile.
The uncertainties in the stellar models begin to matter when the rotation  gradient is about six times greater (for \otto it is when the gradient is about four times greater).
This figure also shows that the RLS inversion does  better as the difference between the core and surface rotation increases. 

\begin{center}
\begin{figure}
\vspace{-1cm}
\hspace{-1cm}
\includegraphics[width=0.5\textwidth]{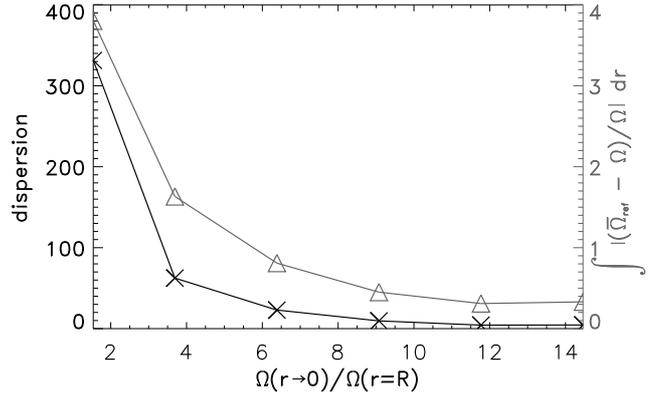}
\caption{
Dispersion ratio values (black crosses, left vertical axis) and the integrated difference between the inverted rotation profile  and the smooth  rotation model  (grey triangles, right vertical axis) as a function of the core-to-surface rotation rate gradient.
}
\label{fig:refbob}
\end{figure}
\end{center}

\section{A sufficient inversion method: Functional fitting}\label{sect:funcfit}
Considering the large observational uncertainties and insensitivity to the stellar models in the radially resolved RLS inversions, it is sufficient to consider fitting for basic parameters of differential rotation such as the average amplitudes and sign of the gradient. 
Functional fitting forces a specific functional fit to the data. 
We choose to fit a step function as the simplest estimate of the rotation above and below the base of the convection zone.

The rotation rate of the star is expressed as 
\begin{equation}
\Omega(r_j) = \sum_{k=1}^K a_k F_k(r_j),
\end{equation}
where $F_k$ is a function with $a_k$ coefficients to be evaluated, and $r_j$ are the set of radial points to evaluate for.

We choose the simplest function
\begin{equation}
     F(r) =\left\{ 
     \begin{array}{l}
         \Omega_\mathrm{1} \quad 0 < r/R \le r_\mathrm{BCZ}\\     
         \Omega_\mathrm{2} \quad \text{$r_\mathrm{BCZ} < r/R \le 1.0$},       
      \end{array} \right.
\end{equation}
where $r_\mathrm{BCZ}$ is the radius of the base of the convection zone, and the uncertainties on the parameters are $\sigma(a_k)=\sqrt{\sum\limits_{i=1}^M \left[  \sigma_i(\delta \omega) c_i(r_j)  \right]^2}$, and $c_{ij}$ are the inversion coefficients. We solve  for the coefficients $\Omega_\mathrm{1}$ and $\Omega_\mathrm{2}$. 

Figure~\ref{fig:boballways} shows inversion results from three different inversion methods: RLS, optimally localised average \citep[OLA; ][]{BackusGilbert1968} inversions, and functional fitting for \bob. 

The functional fit for the smooth rotation profile returns coefficients $\overline{\Omega}_1=-575\pm1176$~nHz and $\overline{\Omega}_2=1535\pm71$~nHz.

It is difficult to distinguish between the synthetic rotation profiles for any of the inversion methods. All results are consistent with solid-body rotation, so that it is still not possible to determine the sign of the radial differential rotation gradient.
To be able to retrieve the sign of the step function correctly, the combined uncertainty on $\sigma(\overline{\Omega}_1) + \sigma(\overline{\Omega}_2)$ needs to be less than the absolute value step function, $|\Delta \Omega |= 350$~nHz, which means reducing $\sigma(\overline{\Omega}_1)$  and $\sigma(\overline{\Omega}_2)$ by a factor of $\approx 4$.
This shows that all inversion techniques suffer similar limitations as RLS inversions. 

\begin{center}
\begin{figure}
\includegraphics[width=0.5\textwidth]{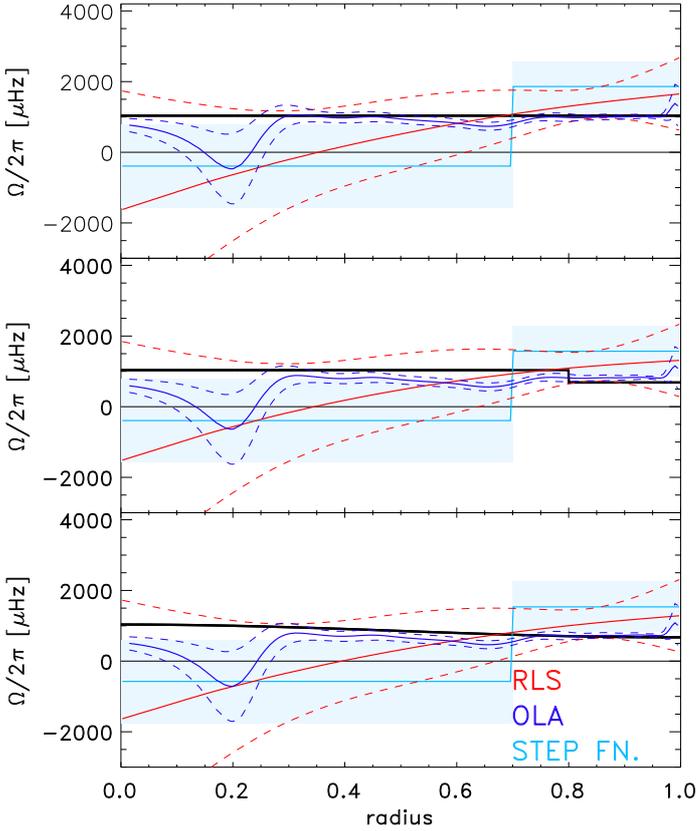}
\caption{
Inverted rotation results for \bob \, and three synthetic rotation profiles, a solid body (top panel), a step function (middle panel), and a smooth function (bottom panel) all shown with a thick black line.
The results from an RLS inversion (red), an OLA inversion (dark blue), and functional fitting of a step function (light blue) are shown in solid lines, with the uncertainty shown with the dashed lines and blue shaded region for the step function.
The uncertainties make it difficult to distinguish a difference in the inversion results for the different synthetic profiles or to exclude solid body rotation.
}
\label{fig:boballways}
\end{figure}
\end{center}

Figure~\ref{fig:ottoallways} shows that the inversion does a particularly good job of constraining the surface rotation rate for the smooth rotation profile, and arguably returns the best estimates of the synthetic rotation profile for all cases.
For \otto, \, the inverted rotation coefficients for the smooth rotation profile are $\overline{\Omega}_1=-743\pm33$~nHz and $\overline{\Omega}_2=163\pm48$~nHz.
Showing that the sign of the radial differential rotation gradient can be constrained, and we can distinguish between a solid body rotation profile and radial differential rotation gradient.

\begin{center}
\begin{figure}
\includegraphics[width=0.5\textwidth]{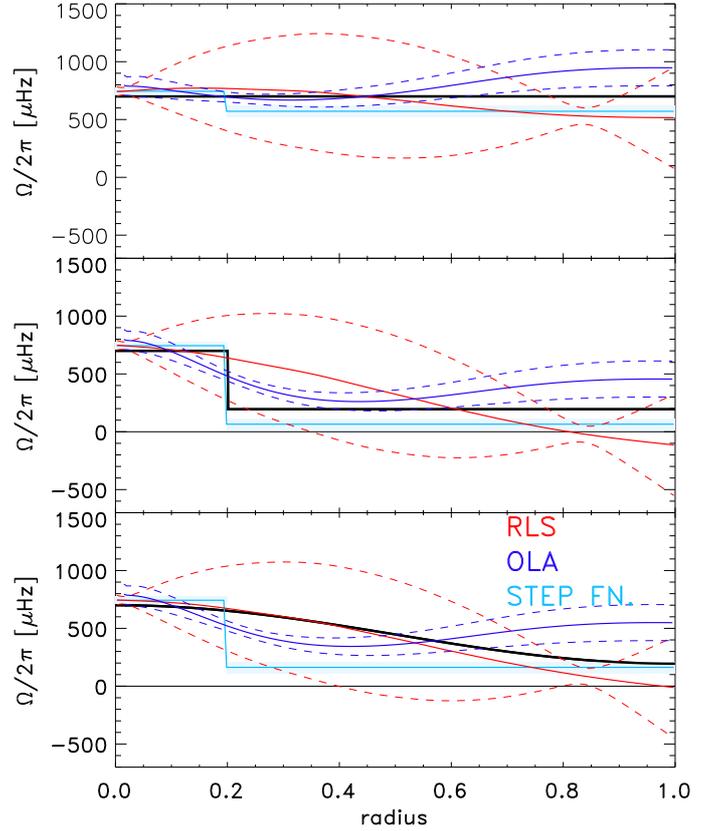}
\caption{
Inverted rotation results for \otto \, with three synthetic rotation profiles: a solid body (top panel), a step function (middle panel), and a smooth function (bottom panel), each shown with a thick black line.
The results from an RLS inversion (red), an OLA inversion (dark blue) and functional fitting of a step function (light blue) are shown in solid lines, with the uncertainty shown with dashed lines.
The uncertainties make it difficult to distinguish a difference in the inversion results for the different synthetic profiles, although the inversion results exclude the solid body case and correctly retrieve the sign of the gradient in the bottom two panels.
}
\label{fig:ottoallways}
\end{figure}
\end{center}

\section{Discussion \& conclusions}\label{sect:conc}

We have demonstrated that for Sun-like stars with moderate radial differential rotation gradients the inversions are insensitive to uncertainties in the stellar models.
The insensitivity to the stellar models for solar-like stars opens the possibility of averaging over many stars to reduce the noise in the splittings.
However, this is not the case for subgiant stars where the sensitivity to uncertainties in the stellar models is comparable to the sensitivity to uncertainties in the rotational splittings.
With improved observations and measurement techniques, one should be careful to use well-constrained stellar models in radial differential rotation inversions for these stars.

We have also demonstrated that it is only possible to extract a few characteristics of the rotation, such as the average magnitude and  gradient, in Sun-like stars.
If a Sun-like star has a core that rotates about six times faster than the surface, then the sign of the rotation  gradient  can be correctly detected. This suggests that the six solar-like stars in \citet{Nielsenetal2014}  have a rotation gradient less than six times that of the Sun.


We did not consider the more optimistic case where, as the rotation rate increases, the splittings become wider than the linewidths, and the splittings are easily measured thereby reducing the uncertainties significantly.

Also, we did not consider the subsequent changes to the stellar models with changes to the asteroseismic constraints,  i.e. variations to the  number of  modes used as constraints or changes in the mode frequency uncertainties. 
It may be worthwhile to investigate at what level of uncertainty in the frequency constraints the spectroscopic uncertainties become important for stellar modelling.

The limitations in the physics of the stellar model should also be kept in mind, with the best fit  of \otto \, having $\chi^2 \approx 20$ (Eqn.~\ref{eqn:chi2}). The systematic uncertainties in the modelling (the input physics, for example, the convective overshoot or effect of rotation on stellar evolution)  may or may not be important for asteroseismic inversions for rotation.



We also found that the radial differential rotation  was unlikely to be well constrained mostly due to the similarity of the sensitivity kernels of the modes. Because of these factors, we argue that radially resolved inversions with regularisation, such as RLS or OLA, are too poorly constrained to use effectively. Therefore, we showed that inverting for a step function rotation profile offers a realistic alternative to infer the average rotation gradient. The method also does surprisingly well compared to the RLS and OLA methods to constrain the surface rotation rate for subgiant star \otto.

Two future observing projects will improve the prospects for characterising the internal rotation of Sun-like stars.
The Stellar Oscillations Network Group \citep{Palleetal2013} is a new ground-based network to measure the surface Doppler velocities of stars, potentially  measuring higher order harmonic degree rotational splittings with smaller magnitudes and  better precision.

The recently selected PLATO mission \citep{Rauer2013} will have two, three-year-long observing runs  of tens of thousands of Sun-like stars. This may reduce the uncertainties in the splittings enough to be able to determine the sign of the radial differential rotation gradient and perhaps the magnitude of the gradient as a function of stellar type.
The insensitivity of the inversions to the solar models suggests that we could exploit the number of main-sequence stars with a broad range of stellar types to infer the  radial differential rotation.
We will explore the  possibility to constrain radial differential rotation using ensemble fits in Sun-like stars in the forthcoming  paper, \citet{Schunkeretal2015b}.

\begin{acknowledgements}
We acknowledge research funding by Deutsche Forschungsgemeinschaft (DFG) under grant SFB 963/1 ``Astrophysical flow instabilities and turbulence'' (Project A18). 
\end{acknowledgements}

\bibliographystyle{aa} 
\bibliography{rotinv} 

\end{document}

%% file: otto_table.tex
\begin{tabular}{ l |   c c c | c | c }
 $\ell$  & 1 & 1,2 & 1,2 & 1,2 & 1,2 \\ 
 $\nu$ [$\mu$Hz]  &   & 285,480 &   & 200,530 & 350,450 \\   
 $\langle \sigma_{i} \rangle~[\mu$Hz] &  93 &  27 &  80 &  99 &  58 \\ 
\hline  \hline \\
 $\mathcal{D}$(solid body)  & 159 & 297 & 165 & 159 & 437 \\ 
 $\mathcal{D}$(step)  & 6.75 & 3.88 & 3.30 & 1.66 & 14.0 \\ 
 $\mathcal{D}$(smooth)  & 6.96 & 6.14 & 3.78 & 2.59 & 19.2 \\ 
\end{tabular}

%% file: bob100_table.tex
\begin{tabular}{ l |  c c c | c c c }
 $\ell$  & 1,2 & 1,2 & 1,2 & 1,2 & 1,2 & 1,3 \\ 
 $n_\mathrm{min},n_\mathrm{max}$  & 16,25 & 16,25 & 16,25 & 10,25 & 18,22 & 16,25 \\ 
 $\langle \sigma_{i} \rangle~[\mu$Hz] &  80 &  8.0 &  1.0 & 140 &  57 &  72 \\ 
\hline  \hline \\
 $\mathcal{D}$(solid body)  & 480 & 43.5 & 21.1 & 357 & 546 & 415 \\ 
 $\mathcal{D}$(step)  & 111 & 23.4 & 18.9 & 115 & 169 & 106 \\ 
 $\mathcal{D}$(smooth)  & 331 & 43.0 & 17.3 & 293 & 416 & 235 \\ 
\end{tabular}